# Controlling asymmetric absorption of metasurfaces via non-Hermitian doping

Min Li[1,2,3], Zuojia Wang[1,2,3*], Wenyan Yin[1,3*], Erping Li[1,3*] and Hongsheng Chen[1,2,3*]

[1]*Interdisciplinary Center for Quantum Information, State Key Laboratory of Modern Optical Instrumentation, College of Information Science and Electronic Engineering, Zhejiang University, Hangzhou 310027, China.*

[2]*ZJU-Hangzhou Global Science and Technology Innovation Center, Key Lab. of Advanced Micro/Nano Electronic Devices & Smart Systems of Zhejiang, Zhejiang University, Hangzhou 310027, China.*

[3]*International Joint Innovation Center, ZJU-UIUC Institute, Zhejiang University, Haining 314400, China*

*Corresponding author: zuojiawang@zju.edu.cn (Z. Wang), hansomchen@zju.edu.cn (H. Chen)

**Abstract**

Metasurfaces based on subwavelength resonators enable novel ways to manipulate the flow of light at optical interfaces. In pursuit of multifunctional or reconfigurable metadevices, efficient tuning of macroscopic performance with little structural/material variation remains a challenge. Here, we put forward the concept of non-Hermitian doping in metasurfaces, showing that an ordinary retroreflector can be switched to asymmetric one by introducing absorptive defects in local regions. The asymmetric absorption performance begins with zero at the Hermitian state, gradually increases under non-Hermitian doping, and reaches the maximum of unity at the exceptional point. This effect is experimentally demonstrated at microwave frequencies via the observation of asymmetric near-field distribution and far-field scattering properties and from a planar metasurface. Furthermore, while importing local gain constituents in conventional retroreflector, it can be tuned to an extremely asymmetric one-side amplifier, extending unidirectional amplification from one-dimensional waveguide to two-dimensional scattering systems. The proposed methodology provides an alternative pathway for engineering electromagnetic metadevices and systems with small perturbations.

**Introduction**

In the past decades, artificially engineered metasurfaces [1-5] have attracted intense attention due to its unique abilities to provide a large degree of control over the amplitude, phase, and polarization of electromagnetic (EM) fields. With judiciously pre-designed nanostructures, metasurfaces have hastened a broad spectrum of novel EM metadevices such as vortex beam generators [6,7], metalenses [8-11], retroreflectors [12-14], holography imagers [15-17] and biosensors [18-20]. To greatly enhance the versatility of metasurfaces and achieve dynamic control over the functions of metasurfaces, substantial efforts have been devoted to equip metasurfaces with various tuning mechanisms. With external stimuli including electrostatic and magnetic forces[21-23], optical or electrical pulses [24,25], voltage bias [26], heat [27,28] and mechanical stretching [29-34], the optical parameters of materials can be elaborately modulated or the structural shapes can be adaptively deformed for tuning functions. Additionally, propagation direction namely *k*-direction has been exploited in the scenario of multifunctional metasurfaces with diverse functionalities. By harnessing judiciously tailored losses, different wave propagations (absorption for one side incidence and anomalous reflection for the other) have been demonstrated [35,36]. So far, the design principle of metasurfaces yields to a continuously transformation philosophy, that is, gradient phase/amplitude distribution provides precise control over the fields. Higher accuracy field manipulation usually requires smoother gradient distribution on constitutive parameters. Similar to transformation optics, the paradigm of continuous gradient has achieved great success and promoted the development of novel electromagnetic devices [1,5,37,38]. However, in multifunctional and reconfigurable scenarios, the philosophy of continuous gradient faces big challenges in engineering. Higher performance requires smoother gradient and thus smaller meta-atoms. Active tuning the electromagnetic responses of enormous meta-atoms one by one could be extremely challenging and large energy-consuming to some extent. Therefore, high-efficiency tuning of macroscopic performance with little structural/material variation is crucial in the design of multifunctional devices.

Recently, photonic doping is emerging as powerful methodology to modulate the macroscopic performance of host media by immersing single or few photonic defects

[39]. Magnetic responses of epsilon-near-zero hosts can be flexibly steered by using only dielectric inclusions. This newly established theory have paved the way for developing flexible nonlinear [40,41], nonreciprocal [42] and quantum metamaterials [43]. Compared with Hermitian photonics, non-Hermitian systems offer a brand new prospective on the role of loss, i.e., losses are not necessarily considered as detrimental effects but can be harnessed to engender anomalous light-matter interactions [44-49]. Examples include unidirectional transmission [45,48,50], negative refraction [47] and exceptional points (EP) [51,52]. Against this background, non-Hermitian doping, in conjunction with parity−time (PT) symmetry in quantum mechanics, enlarges the interplay between gain and loss in the presence of very small gain-loss levels [53].

Here, we put forward the concept of non-Hermitian doping in flat optics and explore complex metasurfaces whose functionalities can be reconfigured by varying local gain or loss distribution. As a proof of concept, we show that an ordinary retroreflector can be switched to asymmetric one by introducing absorptive defects in local regions as illustrated in **Fig. 1**. The asymmetric absorption performance begins with zero at the Hermitian state, gradually increases under non-Hermitian doping, and reaches the maximum of unity at the exceptional point. The asymmetry level can be arbitrarily engineered via tailoring the local loss. This effect is experimentally demonstrated at microwave frequencies via the observation of asymmetric near-field distribution and far-field scattering from a planar metasurface. Furthermore, while importing local gain constituents in conventional retroreflector, it can be tuned to an extremely asymmetric one-side amplifier, extending unidirectional amplification from one-dimensional waveguide to two-dimensional scattering systems.

**Results**

**Theoretical model of asymmetric absorption via non-Hermitian doping.** Let us consider an impenetrable metasurface with periodicity $D_x$ along the $x$ direction, located in the $xy$ plane (see **Fig. 1**). The metasurface is illuminated by TM polarized plane wave propagating in $zx$ plane at an incident angle $\theta_i$. The diffraction mode with the reflection angle $\theta_r$ satisfies [54]:

$$k_0(sin\theta_r - sin\theta_i) = (m+1)G \qquad (1)$$

where $k_0$ is the wave number in free space, $\theta_i$ ($\theta_r$) represents the incident (reflected) angle, $G$ is the amplitude of the reciprocal lattice vector ($G = 2\pi/D_x$, $D_x$ is the period of the metasurface along $x$ direction), and $m$ is the diffraction order. The incident and diffraction angles are defined in the anti-clockwise and clockwise directions with respect to the $z$ axis respectively. $\theta_i$ is positive (negative) for the left- (right-) side incidence. $\theta_r$ is negative (positive) for the left- (right-) side diffraction. The period and the working wavelength of the metasurface are designed to ensure only two propagating diffraction modes, keeping others evanescent. The incident TM polarized wave will be redirected towards two diffraction channels with the reflected angles satisfying Eq. (1): one is anomalous reflection (i.e., retroreflection) and the other is specular reflection. Without loss of generality, the angle of incidence is set to $\theta_i = 45°$ (-45°) for the left- (right-) side incidence. In order to realize retroreflection, the reciprocal lattice vector is set to $G = \sqrt{2}k_0$. For the left-side incidence ($\theta_i = 45°$), as predicted by Eq. (1), only the diffraction orders of $m$ = -1, -2 give rise to propagating waves, corresponding to the specular reflection ($\theta_r$ = 45°) and retroreflection ($\theta_r$ = -45°), respectively. For the right-side incidence ($\theta_i$ = -45°), only the diffraction orders of $m$ = -1, 0 produce propagating waves, representing the specular reflection ($\theta_r$ = -45°) and retroreflection ($\theta_r$ = 45°), respectively. Therefore, such metasurface can be treated as a two-port scattering system, with its scattering matrix expressed by [55]

$$S = \begin{bmatrix} r_{-1}^+ & r_0^- \\ r_{-2}^+ & r_{-1}^- \end{bmatrix} \qquad (2)$$

Here, $r$ is the reflection coefficient of each diffraction mode. The subscript indicates the mode order $m$, and the superscript + (-) indicates the left- (right-)side incidence. The reciprocal theory requires $r_{-1}^- = r_{-1}^+$, indicating identical strength in specular reflection for left- and right-side incidences. However, as shown in the following, the anomalous reflection of left-side incidence $r_{-2}^+$ can be highly different from the right-side one $r_0^-$.

To realize such gradient metasurface, we divide a period structure (covering $2\pi$ range) into $N$ subunits, each of which is generalized by a dielectric slab with an effective refractive index $n_i$ ($i$ = 1, 2, …, $N$), as shown in **Fig. 2**A. The dielectric slab has a thickness $d$ in the $z$ direction, and a width $D_x/N$ in the $x$ direction. Notably, neighboring dielectric slabs are separated by a rigid wall made of perfect electric conductor (PEC). To generate the phase gradient along the surface and cover the $2\pi$ phase range, we use the exact refractive index dictated by the gradient index. An additional term $L$ (any real number) is also introduced to shift the index profile, so that the refractive index distribution reads

$$n_i = (i+L)\beta, i = 1,2,...,N. \qquad (3)$$

with $\beta = 2Nd\lambda^{-1}$ and $\lambda$ is the working wavelength. Next, the lossless phase gradient metasurface is doped with absorptive defects in local regions. For simplification, we assume the metasurface consists of 6 subunits in one supercell, i.e., $N = 6$. Loss is introduced into 1st sub-unit (can be any other subunit) whereas other units are assumed lossless. Therefore, the refractive index of 1st subunit is a complex number $n_1 = n_1^r + in_1^i$. The real part of $n_1$ is expressed by Eq. (3) while the imaginary part can be flexibly customized to explore the asymmetric properties of the metasurface. Then, by using couple mode theory, we can obtain the complete reflection spectrum of the metasurface. The detailed derivations are included in the Supporting Information. **Figs. 2**B-C show the calculated amplitude and phase performance of retroreflection versus the term $L$ and the loss $n_1^i$ of the first subunit for left- and right-side incidences. When there is no loss, the metasurface behaves as a high-efficient retroreflector for both left- and right-side incidences. However, the case is dramatically different when introducing loss into the first sub-unit. An interesting point is that a zero reflectivity and phase singularity can be observed at $L = 7.96$ and $n_1^i / \beta = 0.65$, and this can be identified as an EP of the metasurface. As a contrary, for right-side incidence, retroreflection amplitude remains a high value with the increasing loss in 1st subunit. To make it clearer, **Fig. 2**D plots the retroreflection amplitude against the loss $n_1^i$ with the term $L$ set to 7.96. When the 1st subunit is lossless, near-unitary retroreflection arises from both sides in a symmetric fashion. As the loss initially increases, the amplitude of left-side retroreflection $r_{-2}^+$ declines. A peculiar case where the left-side retroreflection vanishes can be observed at the EP when $n_1^i$ reaches $0.65\beta$. Interestingly, left-side retroreflection $r_{-2}^{+}$ increases once the loss goes beyond the EP. Meanwhile, the right-side retroreflection $r_0^{-}$ is stable against the variation of loss. This indicates the asymmetric ratio can be controlled by tailoring the loss in 1st subunit. We have also performed full wave simulations in commercial software (CST Microwave Studio), the simulation results are represented by solid dots shown in **Fig. 2**D, which fits well with the theoretical results. Notably, considering the edge effects induced by the rigid PEC walls for separating neighboring sub-units, the physical thickness of the metasurface is set to $d = 18.6$ mm in the simulations whereas the geometry thickness $d = 20$ mm in the calculation part. We further calculate the eigenvalues $\lambda_{\pm}$ of the scattering matrix described in Eq. (3). Their trajectories with the loss $n_1^i$ are shown in **Fig. 2**E. Here, the EP is a crossing point of two eigenvalues, which is similar to what has been observed in 1D passive *PT* -symmetric systems [50,56,57]. At the EP, the two eigenvalues $\lambda_{\pm}$ coalesce together with the corresponding eigenvectors transiting from originally orthogonal to parallel relationships. This is a direct manifestation of the EP from the non-Hermiticity of the system [49].

**Numerical and experimental demonstration of asymmetric metasurfaces with largely reduced footprint and doped regions**. The periodic dielectric metasurface is limited inherently by its bulk and rough configuration since the accumulated phase profile requires sufficient propagation distance within the slab. Besides, the introduced losses are isotropically distributed in the local regions. In the following, we focus on non-Hermitian doping in the planar asymmetric retroreflector based on metallic

resonators, dramatically reducing both the footprint of the whole structure and the areas of doped regions. We first apply the resistor as absorptive defects into the local regions shown in **Fig. 3**A. The subunit consists of a dielectric layer (permittivity $\varepsilon = 3$, tangential loss $\delta = 0.003$ at 10 GHz), PEC wall (thickness $t = 0.5$ mm) and a metallic resonator with thickness of 0.035 mm. All the units have the same groove width $p = 10$ mm and depth $d = 5$ mm, one period of the supercell is $D_x = 66$ mm. The resistor (red line) is inserted in the gap ($g = 1$ mm) of the copper resonator. The parameters of the resonators are listed in **Table S1**. With the doping resistance increases, it is expected the absorption efficiency of the loss unit increases, corresponding to the decreased reflection amplitude in **Fig. 3**B. Meanwhile, the phase response is very stable against varied resistance. The reflection amplitude in **Fig. 3**C shows identical trend as shown in **Fig. 2**D. The extremely asymmetric retroreflection occurs when the resistance is 50 Ω with left-side retroreflection amplitude decreases to 3.7% while right-side retroreflection maintains 95.4%.

To validate the theoretical and numerical results, corresponding experiment has been carried out. The first subunit of the fabricated sample is opened with a slit (in order to introduce loss) and is covered by a layer of microwave absorptive material at the end shown in **Fig. 3**E. The length of the slit is $l = 8$ mm, and the width is $w = 2$ mm. The overview of the fabricated sample is illustrated in **Fig. 3**D. The parameters of top metallic resonators are listed in **Table S2**. **Fig. 3**F represents reflection amplitudes and phase response for different subunits. We place absorptive material around the sample to minimize unwanted reflections and the experimental set up is illustrated in **Fig. S2**.

More information about the sample fabrication and experimental measurement are included in **Materials and Methods.** Good agreement can be found between the simulated [**Fig. 3**G] and measured results [**Fig. 3**H], with both suggesting asymmetrical wave behavior. Furthermore, we measured the reflection amplitude towards the specular reflection channel shown in **Fig. S**3, indicating the specular reflection is largely suppressed at working frequency 3.2 GHz. **Fig. 3**I displays the measured far-field scattering of the metasurface, which is normalized by the scattering intensity from PEC with same size. It should be pointed out that, although the EP is achieved at 45° for the current design, strongly asymmetrical behavior is in fact observable under a wide range of angles of incidence even under large ones (e.g. $\theta_i = \pm 80°$) (see **Fig. S**4).

**Extremely asymmetric amplification via non-Hermitian doping.** We further investigate non-Hermitian doping as unique tool to obtain asymmetric amplification while importing active constituents into local regions of the metasurface. Such methodology extends unidirectional amplification from previous one-dimensional waveguide to two-dimensional scattering systems. The schematic view of the asymmetric amplification is shown in **Fig. 4**A. Conventional symmetric retroreflector is doped with the active constituents (red areas) in local regions, which can be reconfigured by varying the optical gain (e.g., via optical pumping). Here, the active constituents are represented by a dialectic slab with negative imaginary part of refractive index. All the other parameters are identical with the loss case discussed above. Theoretical reflection spectrums in **Fig. 4**B indicate an extremely asymmetric amplification at $L$ = 7.94 and $n_1^i/\beta = -0.55$ with $r_{-2}^+ = 185$ and $r_0^- = 14$. The asymmetric performance is also investigated by COMSOL MULTIPHYSICS. **Fig. 4**C shows the simulated results and asymmetric point appears at $L$ = 7.26 and $n_1^i/\beta = -0.56$ with $r_{-2}^+ = 257$ and $r_0^- = 17$. Notably, the deviation of the term $L$ between theoretical and simulated results is $\Delta L$ = 0.68, corresponding tiny deviation of refractive index, specifically, the difference between the theoretical and simulated refractive index of the dielectric slab is $\Delta n = \Delta L * \beta$ = 0.077. The deviation of amplification amplitude mainly comes from the resolution of the discrete term $L$ and introduced gain $n_1^i$ in theoretical and simulated procedures. **Fig. 4**D plots the retroreflection amplitude against the gain $n_1^i$ with the term $L$ set to 7.26. The inset figures show the scattering magnetic field distribution in $zx$ plane, suggesting an asymmetric amplification fashion for left- and right-side incidence.

**Discussion**

Here, we have theoretically and experimentally demonstrated the concept of non-Hermitian doping in the scenario of complex metasurfaces. Macroscopic performance of the metasurface can be efficiently modulated by immersing loss or gain in local regions. As a proof of concept, we show a conventional symmetric retroreflector is switched to an asymmetric absorber (asymmetric amplifier) while the local areas are doped with absorptive defects (active constituents). The asymmetric level can be flexibly controlled by tailoring the local gain or loss. Compared with continuously transformation philosophy-based design of metasurfaces, our methodology provides an alternative pathway for high-efficient engineering electromagnetic metadevices and systems with small perturbations.

**Materials and Methods**

**Microwave sample fabrication.** The samples were firstly fabricated by a standard printing circuit board technology, periodically printed on a F4B dielectric slab (thickness 5 mm, permittivity 3.5 and tangential loss 0.003 at 10 GHz). The thickness of copper layer is 0.035 mm. After laser beam cutting, the long strips were inserted into the periodical PEC grooves (thickness $t = 0.5$ mm). All the units have the same groove width $p = 10$ mm and depth $d = 5$ mm, one period of the supercell is $D_x = 66$ mm. The sample consists of 48*20 unit cells, the whole size of the sample is 0.528 m along $x$ direction and 0.2 m along $y$ direction.

**Experimental measurements.**

**Near-field scanning.** For the measurement, a broad bandwidth antenna (2-20 GHz) was used to illuminate TM-polarized plane wave on the sample, and a dipole antenna to probe the electric field along the $z$ direction was fixed on the 3D near-field scanning system. The transmitter antenna and the probe antenna were connected with a vector network analyzer (VNA). The experimental set up for near-field scanning is shown in **Fig. S2**A. The probe moved with a step-size of 5 mm and whole scanning area was 500*400 mm in $zx$ plane. The scattered field is obtained by subtracting the incident field (scanned without the metasurface) from the total field (scanned with the metasurface).

**Far-field measurement.** The far-field scattering properties of the sample has been experimentally verified in an anechoic chamber with a setup of two wideband antennas connected with a vector network analyzer. The sample composing of 48 × 20 unit cells has been fabricated and put at the center of an arc track with radius of 1.2 m. The two antennas have been mounted on the arc track. One of the horn antennas serves as the source to illuminate oblique incidence plane wave to the sample. Another horn antenna works as the receiver moving along the arc track to measure the scattered fields at various reflection angles. The experimental set up for far-field measurment is shown in **Fig. S2**B.

**Supplementary Materials**

Supplementary information is available for this paper at https://doi.org/xx.xxxx/sxxxxx-xxx-xxxx-x.

**Funding:** The work was sponsored by the National Natural Science Foundation of China (NNSFC) under Grants No. 61801268, No. 61625502, No. 11961141010, No. 61975176), the Top-Notch Young Talents Program of China, the Fundamental Research Funds for the Central Universities. **Author contributions:** All authors contributed extensively to the work presented in this paper. M.L. and Z.W. initiated the idea; M.L. performed the simulation and experiment; Z.W., W.Y., E.L and H.C. analyzed the data and interpreted detailed results; M.L. and Z.W. wrote the manuscript; Z.W. and H.C. supervised the project. **Competing interests:** The authors declare no competing financial interests. **Data availability:** The data that support the plots within this paper and other finding of this study are available from the corresponding author upon reasonable request. Correspondence and requests for materials should be addressed to Z.W. or H.C.

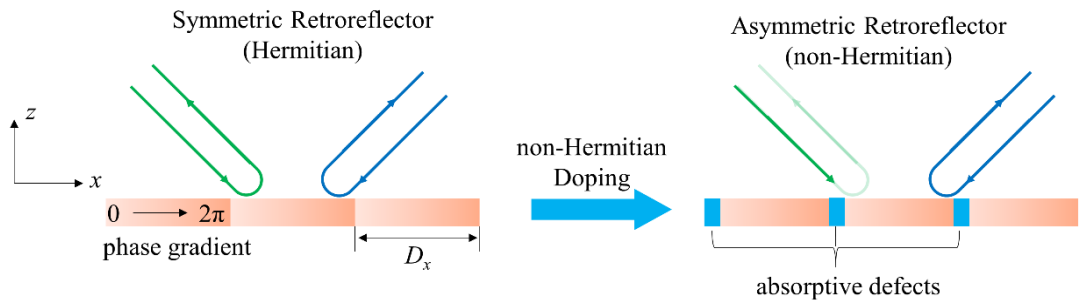

**Fig. 1. Schematic of the non-Hermitian doping for exploring complex metasurfaces.** Left panel represents an ordinary retroreflector made of phase gradient metasurface, showing symmetric high-efficiency retroreflection for left- and right-side incidence. While the metasurface is doped with absorptive defects, it can be tuned to an extremely asymmetric retroreflector at the EP, i.e., near-perfect absorption for left-side incidence and high-efficiency retroreflection for the opposite side.

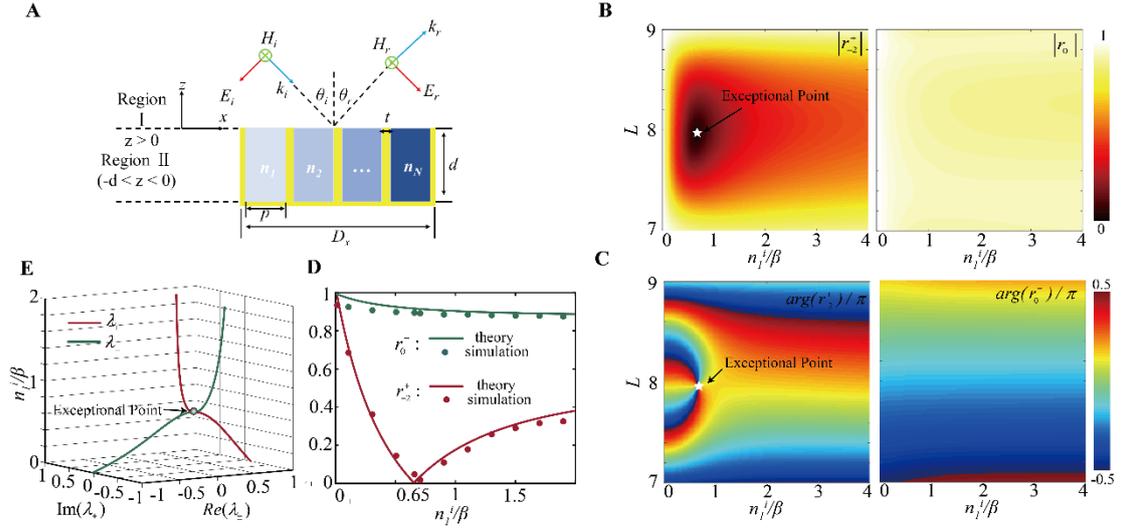

**Fig. 2**. **Theoretical modal of asymmetric absorption via non-Hermitian doping.** (**A**) Schematic of the phase gradient metasurface (covering $2\pi$ phase range) with $N$ subunits, illuminated by TM polarized incidence wave. (**B** and **C**) Calculated retroreflection amplitude spectrum (left panel) and phase spectrum (right panel) for: (**B**) left-side incidence and (**C**) right-side incidence. The exceptional point (marked by white star) is observed at $L = 7.96$ and $n_1^i/\beta = 0.65$. (**D**) Retroreflection amplitude versus loss in first subunit for left- and right-side incidence. (**E**) Trajectories of the eigenvalues of the scattering matrix with the evolution of introduced loss.

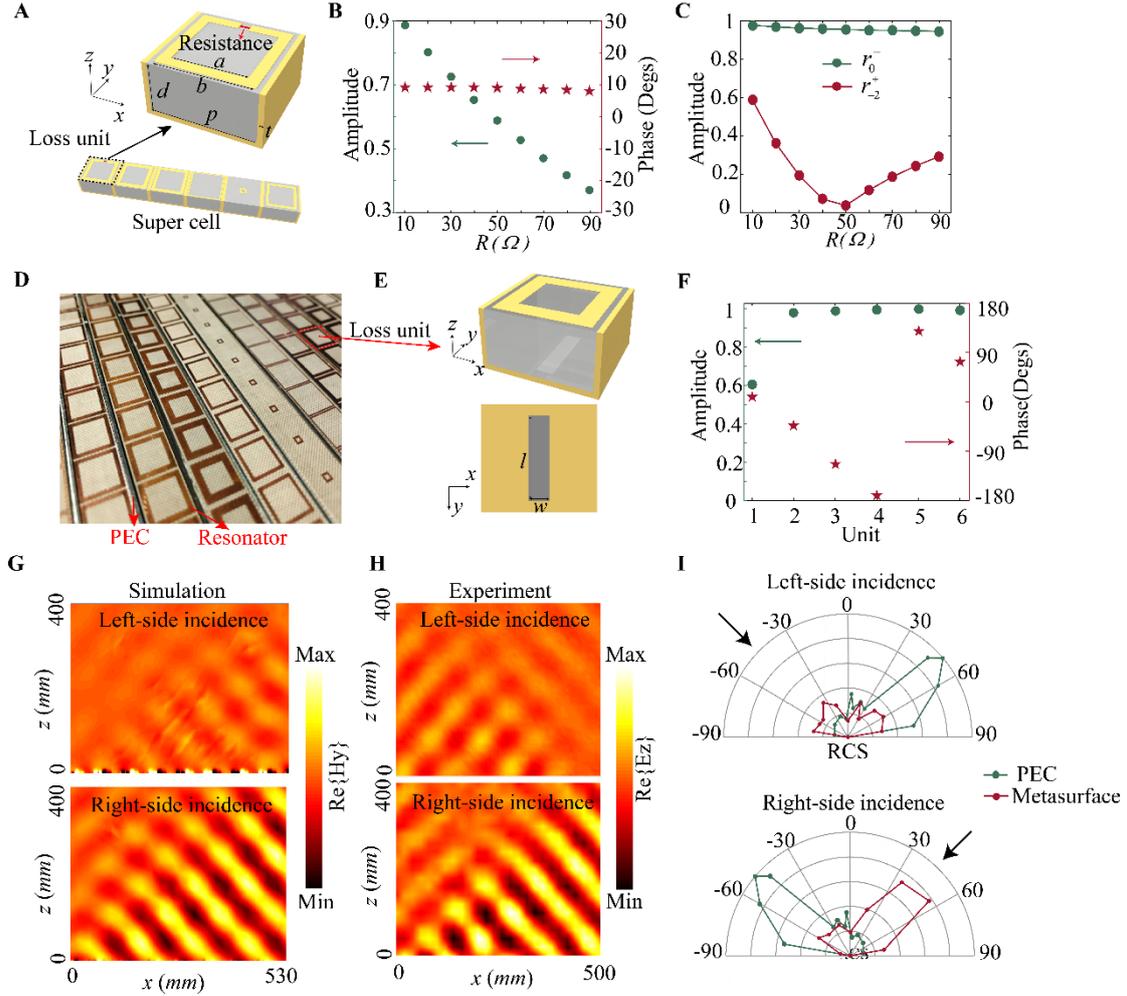

**Fig. 3. Numerical and experimental demonstration of planar asymmetric metasurfaces.** (**A-C**) Numerical calculation for asymmetric retroreflector by immersing resistor as absorptive defects into local regions: (**A**) Details of the supercell and loss unit. (**B**) Reflection amplitudes and phase responses for the loss unit versus resistance value. (**C**) Retroreflection amplitude against resistance for left- and right-side incidence. (**D-I**) Design, fabrication and experiments of a planar asymmetric retroreflector with one subunit opened with a slit for effective loss: (**D**) Overview of the fabricated sample and (**E**) Details of the loss unit. (**F**) Reflection amplitudes and phase responses for different subunits. Near field distribution in *zx* plane with asymmetric fashion for (**G**) Simulated results and (**H**) Experimental results. (**I**) Measured far-field scattering properties of the finite size structure with 48*20 unit cells (0.528 m*0.2 m) for left- and right-side incidence.

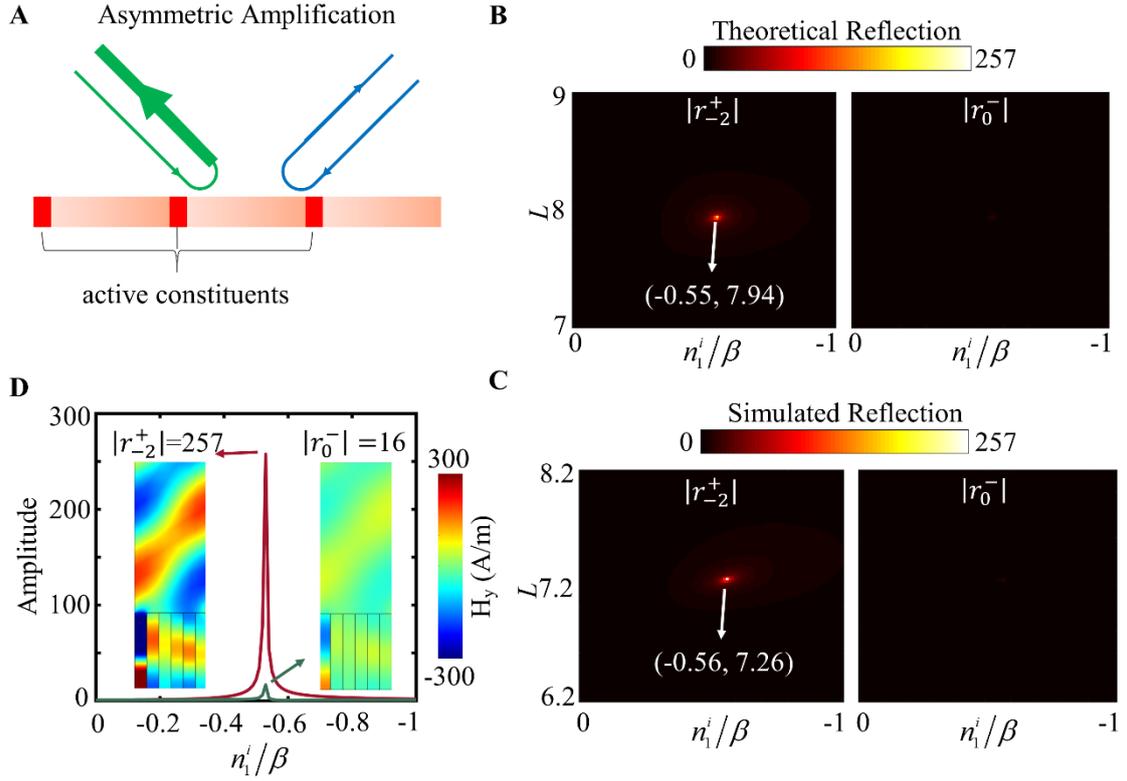

**Fig. 4**. **Extremely asymmetric amplification via non-Hermitian doping**. (**A**) Schematic view of asymmetric amplification while the meatsurface is doped with active constituents (red regions). (**B** and **C**) Reflection spectrum of left- (left panel) and right-side incidence (right panel) versus term $L$ and gain $n_1^i/\beta$ for: (**B**) Theoretical and (**C**) Simulation results. (**D**) Retroreflection amplitude versus gain in first subunit for left- and right-side incidence with the term $L = 7.26$. The inset Figs show the scattering magnetic field distribution in $zx$ plane.

Supplementary Information for

# Controlling asymmetric absorption of metasurfaces via non-Hermitian doping

Min Li, Zuojia Wang, Wenyan Yin, Erping Li and Hongsheng Chen

**Supplementary Information Guide:**

-- Section S1. Derivation of the complete reflection spectrum.

-- Section S2. Parameters of top metallic resonators in planar asymmetric retroreflector.

-- Section S3. Near-field scanning and far-field measurement.

-- Section S4. Wide-angle asymmetric anomalous reflection in the designed planar metasurface.

**Section S1. Derivation of the complete reflection spectrum**

In this section, we use the coupled mode theory as the theoretical model to calculate the reflection coefficient of each diffraction mode.

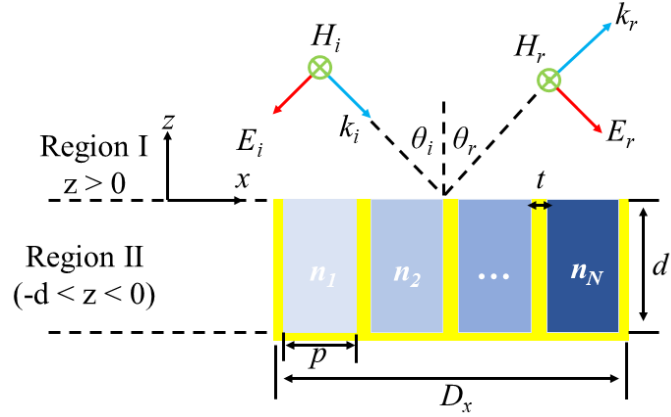

**Fig. S1. Schematic view of the refractive-index gradient metasurface being illuminated by TM-polarized wave with an angle of incidence at $\theta_i$.**

A schematic of a reflective-type metasurface with gradient refractive index is shown in **Fig. S1**. We assume the metasurface is composed of infinite periodical supercells and each supercell consists of $N$ sub-units. In one period, each sub-unit with an effective refractive index $n_i$ (i = 1, 2, ..., N) is separated by PEC (yellow parts). The thickness of the PEC wall is $t$, the width of groove (filled with dielectric slab with refractive index $n_i$) is $p$, and its height is $d$. The period of the metasurface is $D_x$.

The metasurface is illuminated by TM-polarized plane wave propagating in $zx$ plane at an incident angle $\theta_i$. The wavelength of the plane wave is $\lambda$ and time dependence $e^{-i\omega t}$ is assumed. Diffraction mode with reflection angle $\theta_r$ satisfy $k_0(\sin\theta_r - \sin\theta_i) = (m+1)G$. $G$ is the amplitude of the reciprocal lattice vector ($G = 2\pi/D_x$, where $D_x$ is the period of the metasurface along $x$ direction), and $m$ is the diffraction order. The entire domain is divided into two regions, represented in **Fig. S1**. The magnetic field in region Ⅰ ($z > 0$) can be written as

$$H^I = \hat{y}\sum_m (\delta_{-1,m} e^{-ik_{zm}z} + r_m e^{ik_{zm}z})e^{ik_{xm}x} \tag{S1}$$

By solving Maxwell equations, electric field in region I can be written as

$$E^I = -\hat{x}\sum_m \frac{k_{zm}}{\omega\varepsilon_0}(\delta_{-1,m} e^{-ik_{zm}z} - r_m e^{ik_{zm}z})e^{ik_{xm}x} - \hat{z}\sum_m \frac{k_{xm}}{\omega\varepsilon_0}(\delta_{-1,m} e^{-ik_{zm}z} + r_m e^{ik_{zm}z})e^{ik_{xm}x} \tag{S2}$$

where $\delta_{-1,m}$ is the delta function, $\omega$ is the angular frequency, $\varepsilon_0$ represents the permittivity of air in region I, $k_{x,m} = k_0 \sin\theta_i + (m+1)G$, $k_{z,m} = \sqrt{k_0^2 - k_{x,m}^2}$ are z- and x-components of the wave vector of the $m^{th}$ order diffraction mode.

In region II ($-d < z < 0$), the magnetic field inside the $i^{th}$ sub-unit with refractive index $n_i$ is written as

$$H^{II} = \hat{y}(A_i e^{-ik_i z} + B_i e^{ik_i z}) \tag{S3}$$

By solving Maxwell equations, the electric field in region II can be written as

$$E^{II} = -\hat{x}\frac{k_i}{\omega\varepsilon_i}(A_i e^{-ik_i z} - B_i e^{ik_i z}) \tag{S4}$$

where $k_i = k_0 n_i$ is the z-component of the wave vector in the $i^{th}$ sub-unit, $\varepsilon_i$ represents the permittivity of the dielectric slab in the $i^{th}$ sub-unit. Here, we assume the width of each sub-unit is much smaller than the wavelength and ignore the higher modes in the sub-units, so only the fundamental mode in each sub-unit is considered.

On the bottom of the metasurface ($z = -d$), we apply the electric field continuity condition and obtain

$$A_i e^{-ik_i(-d)} - B_i e^{ik_i \cdot -d} = 0 \tag{S5}$$

so that

$$B_i = A_i e^{i2k_i d} \tag{S6}$$

Then by using the magnetic field and electric field continuity condition at the interface of

regions I and II, we obtain

$$\int_{xi-p/2}^{xi+p/2} \sum_m (\delta_{-1,m} + r_m) e^{ik_{xm}x} dx = \int_{xi-p/2}^{xi+p/2} (A_i + B_i) dx \qquad (S7)$$

and

$$\int_0^{D_x} \sum_m \frac{k_{zm}}{\omega\varepsilon_0}(\delta_{-1,m} - r_m) e^{ik_{xm}x} dx = \sum_i^N \int_{xi-p/2}^{xi+p/2} \frac{k_i}{\omega\varepsilon_i}(A_i - B_i) dx \qquad (S8)$$

so that

$$\sum_m (\delta_{-1,m} + r_m) e^{ik_{xm}x_i} sinc(\frac{k_{xm}p}{2}) = (1 + e^{i2k_id}) A_i \qquad (S9)$$

and

$$\frac{k_{zm}}{\varepsilon_0} D_x (\delta_{-1,m} - r_m) = \sum_i^N \frac{k_i p}{\varepsilon_i} e^{-ik_{xm}x_i} sinc(\frac{k_{xm}p}{2})(1 - e^{i2k_id}) A_i \qquad (S10)$$

Here, $x_i$ is the position of middle point of $i^{th}$ sub-unit. **Equations (S9) - (S10)** in fact form a linear equation set where the relative amplitude of the $m^{th}$ order diffraction $r_m$ and $A_i$ remain the unknown values to be solved. For the designed metasurface shown in the main text, the supercell consists of 6 sub-units, i.e., $N = 6$. We consider the diffraction order up to ±100, i.e., $m = 0, ±1, ±2, …, ±100$. By solving the equation set, we can obtain the complete reflection spectrum illustrated in **Fig. 2**.

## Section S2. Parameters of top metallic resonators in planar asymmetric retroreflector

The dimension parameters of metallic resonators of the retroreflector are listed in **Table S1** and **Table S2**.

**Table S1. Parameters list of top metallic resonators in Fig. 3A.**

| i | 1 | 2 | 3 | 4 | 5 | 6 |
|---|---|---|---|---|---|---|

| a (mm) | 3.5 | 3.5 | 3.8 | 4.5 | 0.5 | 3.5 |
| b (mm) | 4.57 | 4.7 | 4.8 | 4.9 | 1 | 4.3 |

**Table S2. Parameters list of top metallic resonators in Fig. 3D.**

| $i$ | 1 | 2 | 3 | 4 | 5 | 6 |
|---|---|---|---|---|---|---|
| a (mm) | 2.8 | 3.5 | 3.8 | 4.5 | 0.5 | 3.5 |
| b (mm) | 4.5 | 4.7 | 4.8 | 4.9 | 1 | 4.3 |

**Section S3. Near-field scanning and far-field measurement.**

In this section, we show the setup of near-field measurement in **Fig. S2**A. A probe is used to detect the electric field, and it can move automatically with a spatial resolution of 5 mm. The far-field measurement setup is shown in **Fig. S2**B. A pair of linearly polarized antennas are fixed on an arch. The measured specular spectra are illustrated in **Fig. S3**. The ordinary specular reflection ($S_{11}$) can be significantly suppressed around the operating frequency.

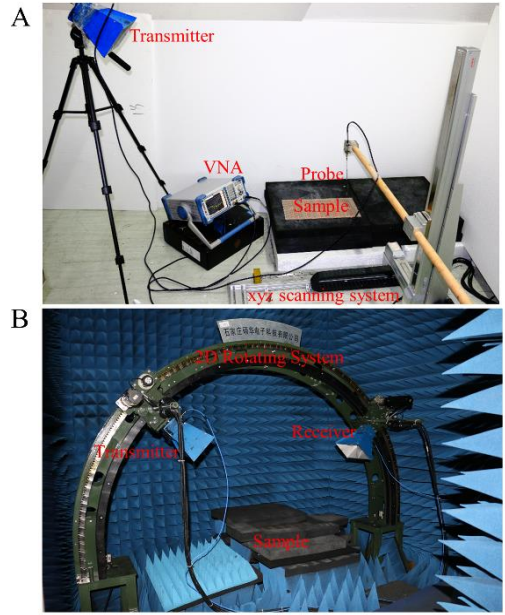

**Fig. S2. Schematic view of experimental setup.** (**A**) Near-field scanning. (**B**) far-field measurement.

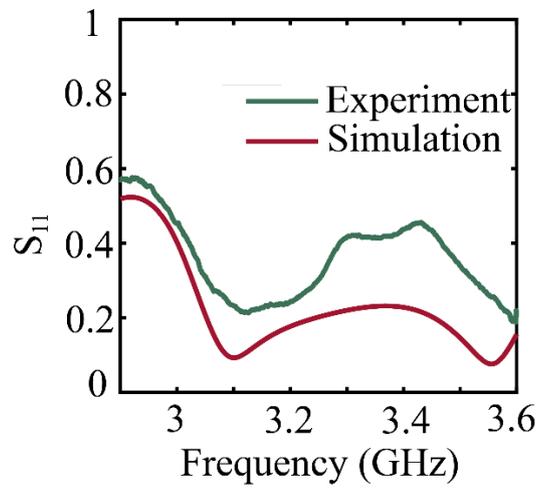

**Fig. S3. Measured specular reflection amplitude.**

**Section S4. Wide-angle asymmetric anomalous reflection in the designed planar metasurface**

In this section, we investigate the performance of planar asymmetric retroreflector shown in

**Fig. 3**D by looking at other angles of the incidence without altering the design. For this case, the anomalous reflection is no longer the exact retroreflection, but it still exists when $|\theta_i| >$ 24.5° according to **Eq. (1)**. **Fig. S4**A shows the anomalous reflection amplitude under opposite incidences, in which a strong contrast can be observed. Notably even the metasurface is illuminated by plane wave under a large incidence angle (e.g. $\theta_i = \pm 80°$), its strong asymmetric responses still retain. We further plot the scattering electric field in *zx* plane with $\theta_i = \pm 30°$ and $\pm 60°$, represented in **Figs. S4**B-C. Figures in the left column shows left-side incidence while the right column shows the right-side incidence. Although the anomalous reflection is still present, it is strongly suppressed compared to the opposite-incidence case, which gives rise to the strong asymmetrical wave behavior. Therefore, the designed planar metasurface can operate under a wide range of angles of incidence.

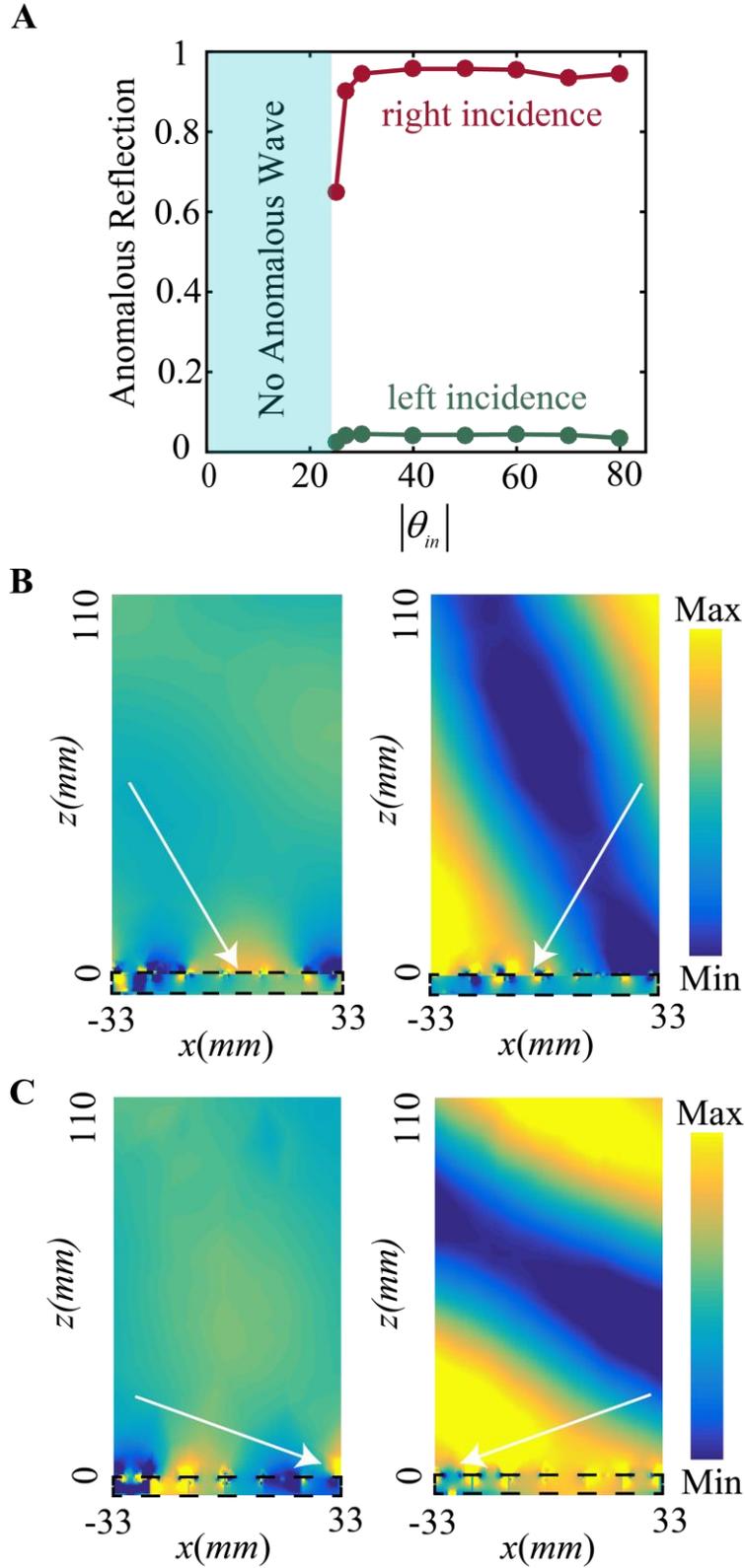

**Fig. S4.** (**A**) Anomalous reflection amplitude for left- and right-side incidence. Scattering electric filed in *zx* plane for: (**B**) $\theta_i = 30°$ (left panel) and $\theta_i = -30°$ (right panel). (**C**) $\theta_i = 60°$ (left panel) and $\theta_i = -60°$ (right panel).